\documentclass[prl,twocolumn,showpacs,preprintnumbers,amsmath,amssymb]{revtex4}
\usepackage{graphicx}
\usepackage{dcolumn,color}
\usepackage{bm}
\usepackage{multirow}
\usepackage{epsfig}
\usepackage[english]{babel}

\begin{document}
\bibliographystyle{try}
\topmargin 0.1cm

\newcounter{univ_counter}
\setcounter{univ_counter} {0}
\addtocounter{univ_counter} {1}
\edef\HISKP{$^{\arabic{univ_counter}}$ } \addtocounter{univ_counter}{1}
\edef\GATCHINA{$^{\arabic{univ_counter}}$ } \addtocounter{univ_counter}{1}
\edef\ERLANGEN{$^{\arabic{univ_counter}}$ } \addtocounter{univ_counter}{1}
\edef\PI{$^{\arabic{univ_counter}}$ } \addtocounter{univ_counter}{1}
\edef\KVI{$^{\arabic{univ_counter}}$ } \addtocounter{univ_counter}{1}
\edef\BOCHUM{$^{\arabic{univ_counter}}$ } \addtocounter{univ_counter}{1}
\edef\DRESDEN{$^{\arabic{univ_counter}}$ } \addtocounter{univ_counter}{1}
\edef\BASEL{$^{\arabic{univ_counter}}$ } \addtocounter{univ_counter}{1}
\edef\GIESSEN{$^{\arabic{univ_counter}}$ } \addtocounter{univ_counter}{1}

\title{Photoproduction of $\boldsymbol\eta$ mesons off protons
for 0.75\,GeV$\,<\,{\mathbf E}_{\boldsymbol\gamma}\,<$\,3\,GeV\\
\vskip 4mm
}
\author{V. Cred\'e~\HISKP$^{\diamond}$,
O.~Bartholomy~\HISKP,
%A.~Anisovich~\mbox{\HISKP\GATCHINA},
A.\,V.~Anisovich~$^{1,2}$,
G.~Anton~\ERLANGEN,
R.~Bantes~\PI,
%D.~Bayadilov~\GATCHINA,
Yu.~Beloglazov~\GATCHINA,
R.~Bogend\"orfer~\ERLANGEN,
R.~Castelijns~\KVI,
%H.~Dutz~\PI,
A.~Ehmanns~\mbox{\HISKP,}
%D.~Elsner~\PI,
J.~Ernst~\HISKP,
I. Fabry~\HISKP,
H.~Flemming~\BOCHUM,
A.~F\"osel~\ERLANGEN,
H.~Freiesleben~\mbox{\DRESDEN,}
M.~Fuchs~\HISKP,
Ch.~Funke~\HISKP,
R.~Gothe~\PI$^+$,
A.~Gridnev~\GATCHINA,
E.~Gutz~\HISKP,
S.\,K.~H\"offgen~\PI,
I.~Horn~\HISKP,
J.~H\"o\ss l~\mbox{\ERLANGEN,}
R.~Joosten~\HISKP~,
J.~Junkersfeld~\HISKP~,
H.~Kalinowsky~\HISKP,
%Frank Klein~\PI,
F.~Klein~\PI,
E.~Klempt~\HISKP,
H.~Koch~\BOCHUM,
M.~Konrad~\mbox{\PI,}
B.~Kopf~\mbox{\BOCHUM\DRESDEN,}
B.~Krusche~\BASEL,
J.~Langheinrich~\PI$^+$,
H.~L\"ohner~\KVI,
I.~Lopatin~\mbox{\GATCHINA,}
J.~Lotz~\HISKP,
H.~Matth\"ay~\BOCHUM,
D.~Menze~\mbox{\PI,}
J.~Messchendorp~\GIESSEN$^\dagger$,
C.~Morales~\PI,
D.~Novinski~\GATCHINA,
%R.~Novotny~\GIESSEN,
M.~Ostrick~\PI,
H.~van~Pee~\mbox{\HISKP$^*$,}
A.~Radkov~\mbox{\GATCHINA,}
J.~Reinnarth~\HISKP,
%S.~Sack~\GIESSEN~,
%A.~Sarantsev~\mbox{\HISKP\GATCHINA},
A.\,V.~Sarantsev~$^{1,2}$,
S.~Schadmand~\GIESSEN,
Ch.~Schmidt~\HISKP,
H.~Schmieden~\PI,
B.~Schoch~\PI,
%A.~S\"ule~\PI,
G.~Suft~\ERLANGEN,
V.~Sumachev~\GATCHINA,
T.~Szczepanek~\HISKP,
U.~Thoma~\HISKP$^*$,
D.~Walther~\PI and
Ch.~Weinheimer~\HISKP \\
(The CB--ELSA Collaboration)}
\affiliation{$^1$ Helmholtz--Institut f\"ur Strahlen-- und Kernphysik,
Universit\"at Bonn, Germany}
\affiliation{\GATCHINA Petersburg
Nuclear Physics Institute, Gatchina, Russia}
\affiliation{\ERLANGEN Physikalisches Institut,
Universit\"at Erlangen, Germany}\affiliation{\PI Physikalisches Institut,
Universit\"at Bonn, Germany}
\affiliation{\KVI Kernfysisch Versneller Instituut, Groningen, Netherlands}
\affiliation{\BOCHUM
Institut f\"ur Experimentalphysik I, Ruhr--Universit\"at Bochum, Germany}
\affiliation{\DRESDEN
Institut f\"ur Kern-- und Teilchenphysik, Universit\"at
Dresden, Germany}
\affiliation{\BASEL Physikalisches Institut, Universit\"at
Basel, Switzerland}
\affiliation{\GIESSEN Physikalisches Institut, Universit\"at
Gie{\ss}en, Germany\\
$^{\diamond}$ currently at Cornell University, USA, $^*$ currently at \GIESSEN, $^\dagger$ currently at \KVI, $^+$ currently at University of South Carolina,USA}

\date{\today}

%-------------abstract----------------

\begin{abstract}

Total and differential cross sections for the reaction p$(\gamma, \eta )$p
have  been measured for photon energies in the range from 750\,MeV to 3\,GeV.
The low--energy data are dominated by the S$_{11}$ wave which has two poles
in the energy region below 2\,GeV. Eleven nucleon resonances are observed in
their decay into p$\eta$. At medium energies we find evidence for a new
resonance  N(2070)D$_{15}$ with
$(M,\,\Gamma) = (2068\pm 22,\,295\pm 40)$\,MeV.
At $\gamma$ energies above
1.5\,GeV, a strong peak in forward direction develops, signalling the
exchange of vector mesons in the $t$ channel.
\end{abstract}

%----------end of abstract-------------

\vskip 5mm

\maketitle

%\section*{Introduction}

\par
Photoproduction experiments provide a sensitive tool to study baryon
resonances.  The information is complementary to experiments with
hadronic beams and gives access to additional properties like helicity
amplitudes.  Baryon resonances have large, overlapping widths
rendering difficult the study of individual states, in particular of
those which are only weakly excited. This problem can be overcome
partly by looking at specific decay channels. The $\eta$ meson
has isospin $I=0$ and consequently, isospin conservation guarantees
that the N$\eta$ final state can only be reached via formation of N$^*$
resonances.  Contributions from $\Delta^{*}$ resonances are excluded.
The $\eta$ meson in the final state thus acts as an isospin filter,
unlike the $\rm\pi N$ channel in which both, $I={1}/{2}$ and
$I={3}/{2}$ states, can appear in the intermediate state. This
selectivity of specific channels is particularly helpful for
coupled--channel analyses. Resonances observed in N$\pi$ or N$\pi\pi$
could belong to the N$^*$ or $\Delta^*$ series, however, even a small
coupling to the N$\eta$ channel identifies them as N$^*$ resonances.
\par
In the near--threshold region, the $\eta$--production process is
strongly dominated by a~single resonance,
N(1535)S$_{11}$~\cite{Krusche:nv}. This resonance has continued to
provoke many theoretical debates due to its unusual parameters. The
branching ratio for N(1535)S$_{11}\rightarrow\eta$N $(\sim 50\%)$ is
much larger than for any other nucleon resonance. As a consequence,
even the very nature of the N(1535)S$_{11}$ as an excited nucleon has
been questioned~\cite{Kaiser:1995cy}. Alternatively, its strong N$\eta$
coupling could be due to mixing between $J=1/2,\,L=1,\,S=3/2$ and
$S=1/2$ quark model states~\cite{IsgurKarl}.  Precise data on the shape
of the N(1535)S$_{11}$ resonance and on its photo--couplings should
help to elucidate its nature. At present, the {\bf P}article {\bf D}ata
{\bf G}roup (PDG) gives a range from 100 to 200\,MeV for its width~\cite{Hagiwara:fs}.  Clearly, high--statistics data covering a wide
range of photon excitation energies are needed to define the properties
of the N(1535)S$_{11}$ more precisely and to identify contributions
from other resonances to the N$\eta$ channel.
\par
In this letter, we present total and differential cross sections for
the reaction $\rm \gamma p \rightarrow p \eta$ as measured with the
CB--ELSA experiment at Bonn covering the entire resonance region and
thus extending the already existing database~\cite{Dytman:vm,Soezueer,Renard:2000iv,Dugger:ft}.
\par
%\section*{Experimental setup}
The experiment was carried out at the tagged photon beam of the {\bf
EL}ectron {\bf S}tretcher {\bf A}ccelerator (ELSA) at the University of
Bonn. The experiment was described briefly in a preceding
letter~\cite{Bartholomy:04} where measurements of total and
differential cross sections for photoproduction of $\pi^0$ mesons
were reported. Also data reconstruction and analysis methods were
described there. The analysis presented here differs
only in the use of two decay modes of the $\eta$ meson,
$\eta\rightarrow 2\gamma$ and $\eta\rightarrow 3\pi^0\rightarrow
6\gamma$.  The acceptance of the detector was determined from
GEANT--based Monte--Carlo simulations. To derive absolute cross
sections the photon flux was used as determined for the reaction $\rm
\gamma p\rightarrow p\pi^0$~\cite{Bartholomy:04}.
\par
\begin{figure}[h!]
\vspace*{-1mm}
\epsfig{file=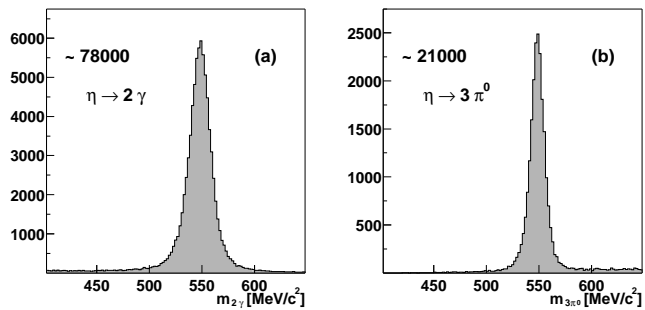,width=0.5\textwidth,clip=}
\vspace*{-3mm}
\caption{(a): The two--$\gamma$ invariant mass spectrum
from events with two photons detected, after a kinematic fit to
$\gamma$p$\rightarrow$p$\gamma\gamma$. (b): The 3$\pi^0$ invariant mass
distribution for events with six detected photons, after fitting
$\gamma$p$\rightarrow$p$3\pi^0$.}
\label{twog} \vspace*{-1mm}
\end{figure}

Fig.~\ref{twog}\,(a) shows the $\gamma\gamma$ invariant mass spectrum
after a $10^{-4}$ confidence--level cut in a kinematic fit enforcing
energy and momentum conservation. Since the proton is identified but
not used as input for the kinematic fit, its momentum is determined by
fitting, resulting in only one kinematic constraint. The $\eta$ meson
is observed above a small residual background.  In a final step, the
mass of the $\eta$ was imposed for the determination of cross sections
in a $\rm\gamma p \rightarrow p\eta\rightarrow p\,2\gamma$
two--constraints kinematic fit (confidence level $>10^{-4}$).
Fig.~\ref{twog}\,(b) shows the  3$\pi^0$ invariant mass spectrum, again
above very little background.  We imposed the pion mass
for three $\gamma\gamma$ pairs in a kinematic fit, resulting in four
constraints. The data were selected by a $10^{-2}$ confidence--level
cut. The residual background events under the $\eta_{\gamma\gamma}$ and
$\eta_{3\pi^0}$ peaks were subtracted using side bins. On average,
there were one to four background events per measured bin.
\par
\par
\begin{figure}
%\vspace*{-6mm}
\hspace{-0.4cm}
\epsfig{file=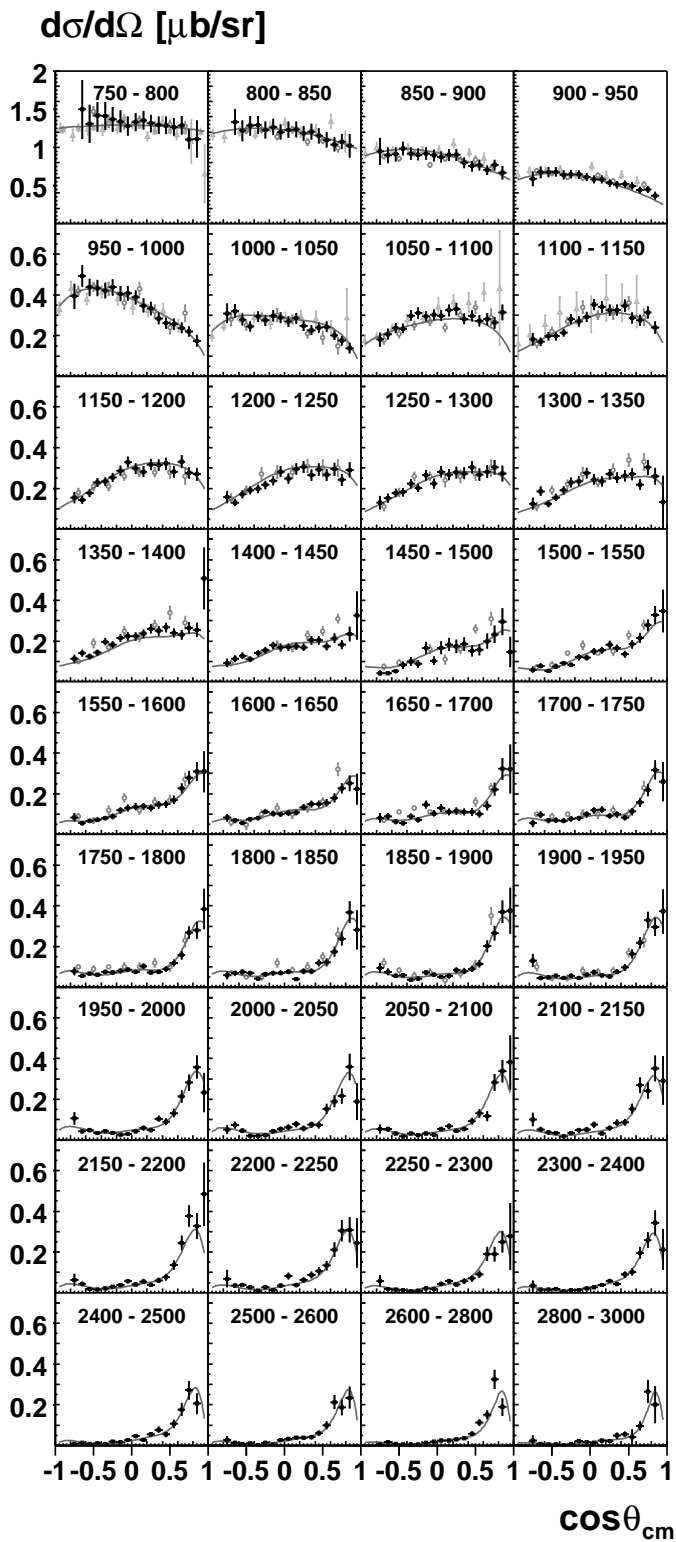,width=0.5\textwidth}
\vspace*{-5mm}
\caption{Differential cross sections for the reaction $\gamma\,\rm{p}
\rightarrow \rm{p}\,\eta$.\\ Experimental data for $E_\gamma = 750$\,MeV to
3000\,MeV: this work (black squares), TAPS~\cite{Krusche:nv},
GRAAL~\cite{Renard:2000iv} and CLAS~\cite{Dugger:ft} data (in
light gray).
The solid line represents the result of our fit.
The error bars represent statistical and systematic errors but
not normalization errors.}
\label{dcs}
\end{figure}

Fig.~\ref{dcs} shows the differential cross sections from both data
sets combined. They are given as functions of the $\eta$ production
angle in the center--of--mass system $\rm \theta_{cm}$ with respect to
the beam axis. The data can be found numerically in~\cite{durham}.
\par
The ratio $\eta\rightarrow 3\pi^0$ over $\eta\rightarrow 2\gamma$ was
determined for each bin in Fig.~\ref{dcs} and histogrammed, giving
$\Gamma_{\eta\rightarrow 3\pi^0}/\Gamma_{\eta\rightarrow 2\gamma} =
0.822\pm 0.002_{\rm stat}\pm 0.004_{\rm syst}$. This value agrees well 
with the PDG value~\cite{Hagiwara:fs} and demonstrates the good
understanding of the detector response. It thus justifies
to add the data from the two channels $\eta\rightarrow 2\gamma$ and
$\eta\rightarrow 3\pi^0\rightarrow 6\gamma$.
\par
The error bars in Fig.~\ref{dcs} represent the statistical and
systematic errors added quadratically.  The systematic errors were
evaluated by changing, in the Monte Carlo simulation, the beam axis
$(=z)$ with respect to the barrel axis by $\pm 3$\,mm, the position of
the target center along  $z$ by  $\pm 1.5$\,mm, and the thickness of
material between target and inner detector by $1$\,mm of capton foil.
These contributions and a relative error of $\pm 5\%$ assigned to the
reconstruction efficiency are added quadratically. An additional error
due to the uncertainty of the normalization is not shown in
Fig.~\ref{dcs}.  It is estimated to be 5\% for photon energies up to
1.3\,GeV and 15\% above as discussed in~\cite{Bartholomy:04}.
\par
\begin{figure}[t]
\vspace*{-1mm}
\epsfig{file=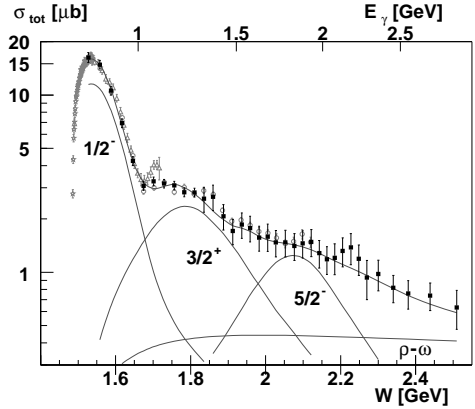,width=0.40\textwidth,clip=}
%\vspace*{-1mm}
\caption{Total cross section (logarithmic scale) for the reaction
$\gamma\,\rm{p}\rightarrow\rm{p}\,\eta$. See caption of Fig.~\ref{dcs}
for symbols. The black squares represent the summation over the angular
bins (bins not covered by measurements are taken from the fit), the
solid line represents our fit. The errors are dominantly due to
uncertainties in the normalization. The contributions of the two
S$_{11}$ resonances, of the $\rm N(1720)P_{13}$, of the $\rm
N(2070)D_{15}$, and of the background amplitudes (mainly $\rho-\omega$
exchange) are shown as well.}
\label{total}
%\vspace*{-2mm}
\end{figure}

The overall consistency with data from GRAAL~\cite{Renard:2000iv} as
well as from CLAS~\cite{Dugger:ft} is very good. At
$\gamma$p invariant mass $W\sim 1716$~MeV/$c^2$, there
is a small discrepancy between those two data sets, particularly
visible in the total cross section (Fig.~\ref{total}). We emphasize
that our data cover a larger solid angle and a wider energy range
and have very little background below the $\eta$ peak.
At a photon energy of about 750\,MeV, our
detection efficiency for events with low--energy protons in the
backward direction (center--of--mass frame) suffers from a large
systematic error.  For photon energies below 1.3\,GeV, the data are
well reproduced by the partial wave analysis SAID~\cite{Arndt:2003zk} and the unitary isobar model MAID~\cite{Chiang:np}.
Above, small deviations of data and SAID show up which become
increasingly important at higher energies.
\par
\par
%\section*{Partial wave analysis}
The data are interpreted in the framework of the approach developed
in~\cite{alexei}. In addition to the data presented here, we include
the Mainz--TAPS data~\cite{Krusche:nv} on $\eta$ photoproduction to
cover the threshold region, beam--asymmetry measurements of $\eta$ and
$\pi^0$ photoproduction~\cite{GRAAL1,SAID1,GRAAL2}, our own data on
$\gamma\,\rm{p}\rightarrow\rm{p}\,\pi^0$~\cite{Bartholomy:04}, and data
on $\rm\gamma p\rightarrow n\pi^+$~\cite{SAID2}. Resonances are
described by relativistic Breit--Wigner amplitudes except for the two
S$_{11}$ resonances at 1535 and 1650\,MeV for which we use a
two--channel K--matrix ($\pi, \eta$). The background is
described by a reggeized $t$--channel $\rho$--$\omega$ exchange and by
nucleon exchange in the $s$ and $u$ channel. The background amplitudes
contribute $\sim 10\%$ to the data shown in Fig.~\ref{total}.
\par
Here we present the results on the p$\eta$ channel.  The $\chi^2$
values for the final partial wave analysis
solution are given in Tab.~\ref{chi}.  The fit
uses 11\, N$^*$ resonances coupling to N$\eta$.  We consider fractional
contributions above 1--2\% as established in this analysis.  
The masses and widths of the observed states are presented in
Tab.~\ref{resonance_list}. We also include the ratio of the helicity
amplitudes $A_{1/2}/A_{3/2}$ and the fractional contribution
normalized to the total cross section for our $\eta$ photoproduction
data. The errors are estimated from a large number of fits in which the
number of resonances, their parameterization, and the relative weight of
the different data sets is changed.
\par
\begin{table}[t]
\caption{Data used in the partial wave analysis
and $\chi^2$ contributions.}
\begin{footnotesize}
\renewcommand{\arraystretch}{1.2}
\begin{tabular}{llcccr}
\hline\hline
 Observable && $N_{\rm data}$   & $\chi^2$ & $\chi^2/N_{\rm data}$   &
 Ref. \\
\hline \hline
$\rm\sigma(\gamma p \rightarrow p\eta)$ & CB--ELSA & 667 & 618 & 0.93 &
~\\ \hline $\rm\sigma(\gamma p \rightarrow p\eta)$ & TAPS & 100 & 160 &
1.60 &
 \cite{Krusche:nv}\\
\hline
$\rm\Sigma(\gamma p \rightarrow p\eta)$ & GRAAL 98 & 51 & 97 & 1.90 &
 \cite{GRAAL2}\\
\hline
$\rm\Sigma(\gamma p \rightarrow p\eta)$ & GRAAL 04 & 100 & 164 & 1.64 &
 \cite{GRAAL1}\\
\hline
$\rm\sigma(\gamma p \rightarrow p\pi^0)$ & CB--ELSA & 1106 & 1750 & 1.58 &
 \cite{Bartholomy:04} \\
\hline
$\rm\Sigma(\gamma p \rightarrow p\pi^0)$ & GRAAL 04 & 359 & 1980 & 5.50 &
 \cite{GRAAL1}\\
\hline
$\rm\Sigma(\gamma p \rightarrow p\pi^0)$ & SAID & 593 & 1470 & 2.48 &
 \cite{SAID1}\\
\hline
$\rm\sigma(\gamma p \rightarrow n\pi^+)$ & SAID & 1583 & 4248 & 2.68 &
 \cite{SAID2}\\
\hline
\hline
\end{tabular}
\label{chi}
\renewcommand{\arraystretch}{1.0}
\end{footnotesize}
%\vspace{-.2cm}
\end{table}

\begin{table}[h]
\caption{Masses, widths,
and resonance couplings obtained in the final partial
wave analysis solution}
\renewcommand{\arraystretch}{1.1}
\begin{tabular}{llrcc}
\hline\hline
\footnotesize{Resonance}& {\footnotesize{ M (MeV)}}   &
{\footnotesize{$\Gamma$ (MeV)}}&
 {\footnotesize{$A_{1/2}/A_{3/2}$}} & Fraction \\
\hline
$\rm N(1520)D_{13}$&$1523\pm 4$&$105^{+6}_{-18}$&$~~0.08\pm 0.10$& 0.020\\
PDG &$1520^{+10}_{-5}$&$120^{+15}_{-10}$   &$-0.14\pm 0.06$      & ~  \\
\hline
$\rm N(1535)S_{11}$$^*$  & $1501\pm 5$&$215\pm 25$&  & \\
PDG           & $1505\pm 10$&  $170\pm 80$     &   ~      & \multirow{2}*{0.430}\\
$\rm N(1650)S_{11}$$^*$ &$1610\pm 10$&$190\pm 20$&  &    \\
PDG              & $1660\pm 20$      & $160\pm 10$ &         & \\
\hline
$\rm N(1675)D_{15}$&$1690\pm 12$&$125\pm 20$    &$~~0.06\pm 0.18$& 0.001 \\
PDG & $1675^{+10}_{-5}$ & $150^{+30}_{-10}$& $1.27\pm0.93$& \\
\hline
$\rm N(1680)F_{15}$&$1669\pm 6 $&$ 85\pm 10$    &$-0.12\pm 0.04$ & 0.005 \\
PDG           & $1680^{+10}_{-5}$ & $130\pm 10$ & $-0.11\pm 0.05$ & ~\\
\hline
$\rm N(1700)D_{13}$&$1740\pm 12$&$ 84\pm 16$    &$~~0.01\pm 0.20$& 0.004 \\
PDG & $1700\pm 50$ & $100\pm 50$ & $9.00\pm 6.5$& \\
\hline
$\rm N(1720)P_{13}$&$1775\pm 18$&$325\pm 25$    &$~~0.68\pm 0.10$& 0.300 \\
PDG           & $1720^{+30}_{-70}$& $250\pm 50$ & $-0.9\pm 1.8$ & ~\\
\hline
$\rm N(2000)F_{15}$&$1950\pm 25$&$230\pm 45$    &$~~1.08\pm 0.60$& 0.007 \\
\hline
$\rm N(2070)D_{15}$&$2068\pm 22$&$295\pm 40$    &$~~1.37\pm 0.24$& 0.171 \\
\hline
$\rm N(2080)D_{13}$&$1943\pm 17$&$ 82\pm 20$    &$~~0.97\pm 0.28$& 0.011 \\
\hline
$\rm N(2200)P_{13}$&$2214\pm 28$&$360\pm 55$    &$~~0.41\pm 0.22$& 0.051  \\
\hline\hline
\end{tabular}\\
\footnotesize{$^*$ K matrix fit, pole position
of the scattering amplitude in the complex plane, fraction for the total K--matrix contribution}\\
\renewcommand{\arraystretch}{1.0}
\label{resonance_list}
\end{table}
Omitting the $\rm N(2070)D_{15}$
changes the $\chi^2$ by 200 for the data of Fig.~\ref{dcs}.
Replacing the $J^P$ assignment from $5/2^-$ to $1/2^\pm$,~...,
$9/2^\pm$, the $\chi^2$ deteriorates by more than 100.
The closest description was obtained fitting with a $9/2^+$ state. For this $J^P$ assignment, the $\chi^2$ was worse by 106 for our data
and by 96 for the beam asymmetry data
$\rm\Sigma(\gamma p\rightarrow p\eta)$~\cite{GRAAL1}.  However, the description of
our $\rm\gamma p\rightarrow p\pi^0$ data deteriorated by more than 500.
We stress that $\rm N(2070)D_{15}$ was required in the analysis
of the data of Fig.~\ref{dcs} alone.
\par
Omitting the $\rm N(2200)P_{13}$ from the analysis changes the $\chi^2$
by 65 for our data and by 45 for the beam--asymmetry data. This
resonance is less significant than the $\rm N(2070)D_{15}$, and
replacing it with a $7/2^+$ state makes the description of our data
worse by only 24.  However, the $\rm N(2200)P_{13}$ is still the
preferred solution for the description of the data in the high mass
region.  We do not find evidence for a third S$_{11}$ for which claims have been
reported at masses of 1780\,MeV~\cite{Saghai:2003ch} and 1846\,MeV~\cite{Chen:2002mn}.
\par
The differential cross sections were integrated to determine the total
cross section (Fig.~\ref{total}). The extrapolation to forward and
backward angles uses the result of the partial wave analysis.  The
solid line represents the integration of the  partial wave solution.
\par
\par
\begin{figure}
%\vspace*{-1mm}
\epsfig{file=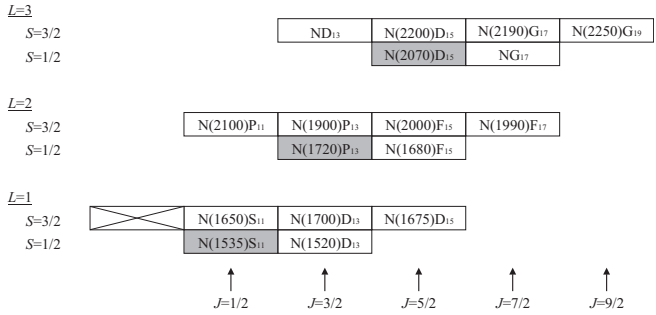,width=0.48\textwidth,clip=}
%\vspace*{-2mm}
\caption{N$^*$ resonances with quantum numbers which can be assigned to
orbital angular momentum excitations with $L=1,2,3$ and with a total
quark spin $S=1/2$ and $S=3/2$. Note that mixing between states of the
same parity and total angular momentum is possible. Resonances with
strong coupling to the N$\eta$ channel are marked in grey.}
\label{resonances}
%\vspace*{-2mm}
\end{figure}

The most prominent contributions to the total cross section stem from a
series of resonances with $\rm N(1535)S_{11}$, $\rm N(1720)P_{13}$, and
$\rm N(2070)D_{15}$ and smaller contributions from $\rm N(1650)S_{11}$
and $\rm N(2200)P_{13}$.  The dominant contributions hence come from
nucleon resonances to which, in the non--relativistic quark model,
quantum numbers $S=1/2$ and $J=L-S$ (Fig.~\ref{resonances}) could be
assigned and which decay into N$\eta$ in a relative S, P, and D wave. 
We do not have an interpretation for this 
observation but obviously, a common explanation of these findings 
is required.
\par
In summary we have reported a measurement of the total and differential
cross sections for the photoproduction of $\eta$ mesons off protons
over a wider range, in energy and in production angle, than covered by
previously
existing data. An isobar analysis of the data which
additionally includes other data sets determines N$\eta$
couplings of eleven N$^*$ resonances, uncovers evidence for
$\rm N(2070)D_{15}$ and gives an indication for
$\rm N(2200)P_{13}$.
\par
We thank
the technical staff at ELSA
and at all the participating institutions for their invaluable
contributions to the success of the experiment. We acknowledge
financial support from the Deutsche Forschungsgemeinschaft (DFG).  The
collaboration with St. Petersburg received funds from DFG and the
Russian Foundation for Basic Research.
\mbox{B.~Krusche} acknowledges
support from Schweizerischer Nationalfond. U.~Thoma thanks for an Emmy
Noether grant from the DFG. A.\,V.~Anisovich and A.\,V.~Sarantsev
acknowledge support from the Alexander von Humboldt Foundation. This
work comprises part of the PhD thesis of O.~Bartholomy.

\end{document}